\documentclass[conference, a4paper]{IEEEtran}
\usepackage{graphicx}
\usepackage{amsmath,amssymb,epsfig,xcolor} 
\usepackage{subfig}
\usepackage{tabularx,multirow,array}
\usepackage{times}
\usepackage{booktabs}
\usepackage{multirow}
\usepackage{soul}
\usepackage{enumerate}
\usepackage{xcolor}
\usepackage[countmax]{subfloat}
\usepackage{algpseudocode}
\usepackage{algorithm}
\usepackage{threeparttable}
\usepackage{cite}
\usepackage{url}
\usepackage{amsthm}
\usepackage{bbm}
\theoremstyle{plain}

 \newtheorem{defn}{Definition}

\SetMathAlphabet{\mathcal}{bold}{OMS}{cmsy}{b}{n}
\begin{document}
\title{Benchmarking Machine Learning Techniques for THz Channel Estimation Problems}
\author{
    \IEEEauthorblockN{Mounir Bensalem  and Admela~Jukan}
    \IEEEauthorblockA{Technische Universit\"at Braunschweig, Germany}
     \IEEEauthorblockA{\{mounir.bensalem, a.jukan\}@tu-bs.de}
}
\maketitle
\begin{abstract}
Terahertz communication is one of the most promising wireless communication technologies for 6G generation and beyond. For THz systems to be practically adopted, channel estimation is one of the key issues. We  consider the problem of channel modeling and estimation with deterministic channel propagation and the related physical characteristics of THz bands, and benchmark various machine learning algorithms to estimate THz channel, including neural networks (NN), logistic regression (LR), and projected gradient ascent (PGA). Numerical results show that PGA algorithm yields the most promising performance at SNR=0 dB with NMSE of -12.8 dB.
\end{abstract}

\section{Introduction}
THz communication systems are positioned as the main driver in the evolution towards next generation wireless communications, also referred to as 6G \cite{saad2019vision}. In THz systems, operating in a frequency range 300 GHz - 10 THz and enabling unprecedented wireless transmission rates, several major physical limitations make the channel modeling and estimation fundamentally hard \cite{petrov2015interference}. Recent work focused on developing systems and algorithms for estimating mmWave channels in MIMO systems \cite{mo2017channel, he2018deep}. MIMO channel estimation is challenging due to the quantization of the linear combination of transmitted signals from multiple input antennas \cite{zhang2018low}. Using higher frequencies, and larger bandwidths needs high precision analog-to-digital-converters (ADCs), which motivates the utilization of low resolution ADCs for MIMO systems \cite{mo2017channel}, and more practically 1-bit ADCs \cite{myers2020low}. 

Today, significant efforts are underway to accurately model and estimate the channel with machine learning (ML). The trend of applying machine learning algorithms in order to estimate mmWave and furthermore THZ channels is rather notable in fact, also considering low resolution ADCs \cite{takeda2019mimo, mo2017channel,chen2020deep, myers2020low}.  \cite{chen2020deep} proposed a deep convolutional neural network (DCNN)-based spherical-wave CE algorithms  for THz systems, the THz channel is simulated using a channel model developed for mmWave systems and adapting the frequency parameters to 0.6 THz. 
We refer to a some ML algorithms studied in \cite{myers2020low} for MIMO systems, such as projected gradient ascent (PGA) and Franke-Wolfe techniques, to benchmark their performance in a THz scenario alongside with basic neural network estimator, due to their good performance for a carrier frequency of 28 GHz.   Overall, ML-based channel estimation problem in THz communication systems is still a rather open research direction, with experimental setups and analytical models still few and far between. 


In this paper, we benchmark different machine learning algorithms used for THz channel estimation. Our reference model is THz-MIMO channel with 1-bit ADCs, as proposed in \cite{lozano20211} and \cite{saad2020mimo} due to the limited quantization level of ultra-high speed low power ADCs. We model the THz channel using a deterministic channel propagation model from the literature \cite{moldovan2014and}, considering the existence molecular absorption loss and showing that data rates around 1 Tbps can be obtained with a short distance of around 1 meter. Our model advances the basic model of THz band proposed in \cite{jornet2011channel}, which uses an updated molecular High Resolution Transmission (HITRAN) 2012 database. We adopt an indoor scenario with reflecting walls in order to consider the line-of-sight (LoS) and several non-line-of-sight (NLoS) reflected rays. We apply the analytical model to obtain the channel state information. Using a pilot generation methods,  Discrete  Fourier Transform (DFT) and Zadoff-Chu (ZC) based sequence training,  we generate a training dataset of pilots input and signal response at the receiver. 
Based on this reference model, we apply different machine learning algorithms for channel estimation, including neural networks (NN), logistic regression (LR), and PGA 
 We chose these specific algorithms  as they proved good performance for similar problems, with complexity similar to the complexity of gradient descent learning 
Numerical results, including analysis and simulations, show that PGA algorithm yields the most promising performance at SNR=0 dB with NMSE of -12.8 dB.

The rest of the paper is organize as follows: 
Section \ref{sec:SM} describes the channel and system model. Section  \ref{sec:ML} describes the studied ML algorithms for Channel Estimation. Section \ref{sec:results} shows the simulation results. We conclude the paper in Section \ref{sec:conclusion}.

\section{Channel and System Model}\label{sec:SM}

\indent \textbf{Notation:} \textbf{H} is a complex matrix and H denotes a real matrix. $\textbf{H}^T$ and $\textbf{H}^\ast$ represent the transpose and conjugate transpose of \textbf{H}, respectively. Re(\textbf{H}), Im(\textbf{H}) are the real and imaginary part of \textbf{H}. $\Vert \textbf{H} \Vert $ is  the Frobenius norm of \textbf{H}. $\mathbbm{1}_{[.]}$ is the indicator function.

\subsection{THz Propagation Model}

We consider a single transmitter (Tx) and a single receiver (Rx) THz transmission system, using a point-to-point line-of-sight communication scenario with omnidirectional antennas \cite{jornet2011channel}, separated by a distance $d$. The main difference between Thz bands and other frequency bands appears from the molecular absorption loss, which varies with  the signal frequency, the transmission distance and the concentration and the mixture of molecules coming across the path. \\
We consider the medium loss due to the vibration changes of molecules, which is denoted as the absorption loss $L_{abs}(f,d)$. This loss accurs due the transformation of a part of the wave energy into a kinetic energy, and it depends on the operating frequency the distance between the sender and receiver, and the composition of the meduim. The equation is obtained from \cite{jornet2011channel} as follows: 
\begin{equation}
L_{abs}(f,d) =  e^{k(f)\cdot d } = k(f)\cdot d \cdot 10 \cdot log(e) \;\;(dB)
\end{equation}
Where f is the operating frequency, and $k(f)$ is is the overall absorption coefficient of the medium available from HITRAN database \cite{rothman2009hitran}, which includes HITRAN line-transition parameters  and cross-sections, that can be used to model signal transmission.

Similar to \cite{jornet2011channel}, we assume that the spreading loss, which represents the attenuation due to the expansion of a wave that  propagates through the medium, is given by:
\begin{equation}
L_{spread}(f,d) = \left ( \frac{4\pi f d}{c} \right )   ^2 = 20 \cdot log \left ( \frac{4\pi f d}{c} \right ) \;\;(dB)
\end{equation}
Where  c is the speed of light  in the vacuum.
The received signal  power spectral densit (psd) in THz  band is given by:
\begin{equation}
\begin{split}\label{eq:powerthz}
P_{Rx} &= \frac{P_{Tx}}{L_{abs}\cdot L_{spread}}\\
& = P_{Tx} \cdot C \cdot f^{-2} \cdot d^{-2} e^{-k(f) d}, \;\;\; C= \frac{c^2 }{16 \pi^2 } 
\end{split}
\end{equation}
Where $P_{Tx}$ represents  the transmitted signal psd. \\
Molecular absorption introduces noise to the transmitted signal, which needs to be considered to evaluate the signal to noise ratio (SNR). The  molecular noise psd is given by \cite{jornet2011channel}:
\begin{equation}\label{eq:molec}
P_M = \frac{P_{Tx}(f)}{L_{spread}(f,d)}(1-e^{-k(f)\cdot d})
\end{equation}
Additionally, we consider the  Johnson-Nyquist (JN) noise generated by thermal agitation of electrons in conductors.  The psd of JN noise is given by\cite{petrov2015interference}:
\begin{equation}\label{eq:njnoise}
P_{JN} = \frac{hf }{ (\exp(\frac{hf}{k_B T}) - 1)}
\end{equation}
Where $k_B$ denotes the Boltzmann constant and $T$ denotes the temperature in Kelvin.\\
In order to simplify our analysis, we  approximate eq. (\ref{eq:njnoise}) using series expansion, then we keep the first order term as follows:
\begin{equation}\label{eq:njapprox}
\begin{split}
P_{JN} & =  \frac{hf }{ (\exp(\frac{hf}{k_B T}) - 1)}\\
      & =  \frac{hf }{\frac{hf}{k_B T} + \left (  \frac{hf}{k_B T}\right )   ^2 \cdot \frac{1}{2!} + \left (  \frac{hf}{k_B T}\right )   ^3 \cdot \frac{1}{3!}+ .. }\\
      & \approx k_B T, \;\;\; \text{1$^{\text{st}}$ order approx.}
\end{split}
\end{equation}
Using eq. (\ref{eq:molec}) and (\ref{eq:njapprox}), the total noise psd is given by:
\begin{equation}\label{eq:noise}
P_N = k_B T +  P_{Tx} \cdot C \cdot f^{-2} \cdot d^{-2}(1- e^{-k(f) d})
\end{equation}
Therefore, the  signal-to-noise ratio (SNR) at the receiver is represented as :
\begin{equation}
\begin{split}
SNR  &= \frac{P_{Rx}}{P_N} 
\end{split}
\end{equation} 
\subsection{System Model}
\begin{figure*}
 \centering
   \includegraphics[scale=1]{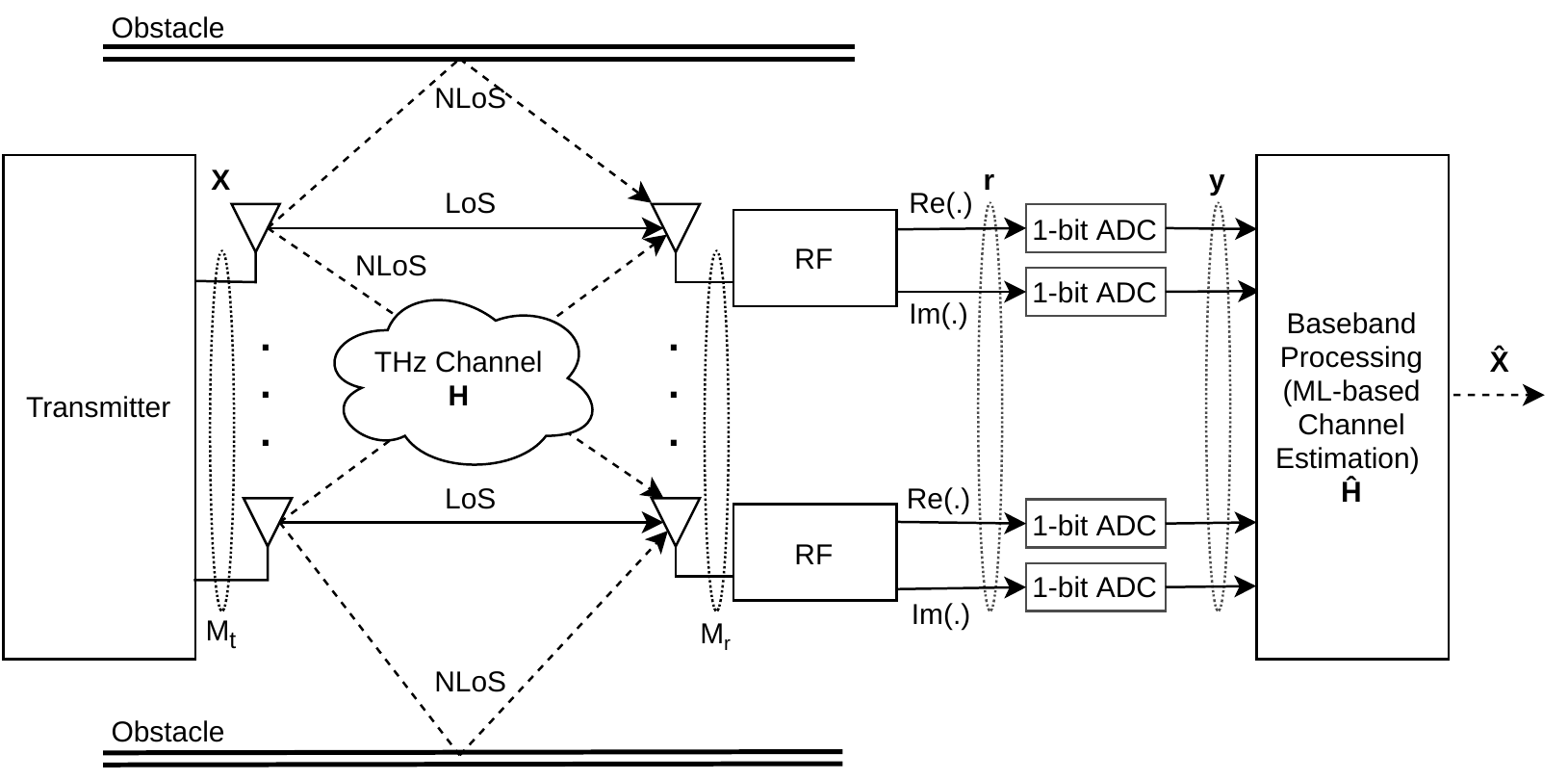}
 \caption{THz-MIMO with $M_t$ transmit antennas, $M_r$ receiver antennas. Two 1-bit ADCs are used to equalize each received signal.}
\label{fig:sysarch}
\end{figure*}

We consider a narrowband THz-MIMO system with $M_t$ antennas at the transmitter, $M_r$ antennas at the receiver. 
Our studied system is depicted in Fig. \ref{fig:sysarch}.
 We assume that the number of antennas at both transmitter and receiver are equal.  The arrival paths are considered to include  two main components of a THz channel:  the LoS and several NLoS reflected rays caused by reflection. 
We assume that the scattering and diffraction can be neglected due to their negligible effect on propagation in the THz band. 

 Let $\textbf{H} \in \mathbb{C}^{M_t \times M_r},$ denotes the narrowband channel  matrix of THz channel impulse response. The channel matrix is expressed as follows \cite{moldovan2014and}:
 \begin{equation}\label{eq:h}
 \begin{split}
 \textbf{H} =& \alpha_L G_t G_r a_r(\theta^r, \psi^r) a_t(\theta^t, \psi^t)\\ & + \sum_{i=1}^{N_{clu}}\sum_{l=1}^{L_{ray}^i} \alpha_{i,l}^{NL} G_t G_r a_r(\theta_{i,l}^r, \psi_{i,l}^r) a_t(\theta_{i,l}^t, \psi_{i,l}^t)P_r(f,\tau_{i,l})
 \end{split}
 \end{equation}
Where $\alpha_L$ is the complex gains of the LOS ray component, given by $|\alpha_L|^2 = L_{spread}(f,d)L_{abs}(f,d) $.  $\alpha_{i,l}^{NL}$
is the complex gains of the NLOS ray component, given by $|\alpha_{i,l}^{NL}|^2 = \mathcal{F}_{i,l}(f) L_{spread}(f,d_1+d_2)L_{abs}(f,d_1+d_2) $, with  $\mathcal{F}_{i,l}$ is the Fresnel reflection coefficient, $d_1$  is the distance between the transmitter and the reflector, and $d_2$ is the distance between the reflector and the receiver \cite{lin2015indoor}. $P_r(\tau_{i,l})$ is the pulse-shaping function at time delay $\tau_{i,l}$.   $G_t$ and $G_r$ are the associated transmit and receive
antenna gains, respectively, $a_t(.)$ and $a_r(.)$  are the antenna array response vectors at the transmitter and receiver, $\theta^t$ , $\psi^t$ , $\theta^r$  and $\psi^r$  are the azimuth AoD, elevation AoD, azimuth AoA, and elevation AoA of the LOS ray component, $\theta_{i,l}^t$ , $\psi_{i,l}^t$ , $\theta_{i,l}^r$  and $\psi_{i,l}^r$  are the corresponding parameters for the $l^th$ NLOS ray in the $i^th$ cluster.

 At the receiver, after obtaining the signal the FFT is computed. Let $\textbf{r} \in \mathbb{C}^{M_r\times 1}$ denote the unquantized received signal, and given as follows:
\begin{equation}
\textbf{r} = \textbf{H} \textbf{x} + \textbf{z}, 
\end{equation} 
where $\textbf{z} \in \mathbb{C}^{M_r\times 1}$ is the  additive  white  gaussian noise   (AWGN) matrix,   whose   samples   are   independent and  identically  distributed  (i.i.d.) $\textbf{z} \sim \mathcal{CN}(0,N_0) $, and $N_0$ is the noise power.\\

\begin{defn} 
An ADC is a device that applies a binary operation on a complex space by mapping its elements into binary numbers. We define the following mappings:
\begin{itemize}
\item 1-bit ADC mapping:
\begin{equation}
\begin{split}
\text{sign}: \mathbb{R} & \longrightarrow  \mathbb{Z}\\ 
x &\longmapsto  \begin{cases}
  1 & \text{if $x\geq 0$ } \\
  -1 & \text{otherwise}
\end{cases},
\end{split}
\end{equation}

\item Vector ADC mapping:
\begin{equation}
\begin{split}
\text{Sign}: \mathbb{R}^{n} & \longrightarrow  \mathbb{Z}^{n}\\ 
(x_1,...,x_n) &\longmapsto  ( \text{sign}(x_1),..., \text{sign}(x_n)),
\end{split}
\end{equation}

\item Matrix ADC mapping:
\begin{equation}
\begin{split}
\text{Sign}: \mathbb{R}^{n\times m} & \longrightarrow  \mathbb{Z}^{n \times m}\\ 
\{x_{i,j}\}_{\substack{1\leq i\leq n\\1\leq j\leq m}} &\longmapsto  \{ \text{Sign}(x_{i,j})\}_{\substack{1\leq i\leq n\\1\leq j\leq m}},
\end{split}
\end{equation} 
\item Element-wise ADC quantization operator for channel response:
\begin{equation}
\begin{split}
\widehat{g}: \mathbb{C}^{n\times m} &\longrightarrow  \mathbb{Z}[j]^{n\times m}\\ 
a+j b & \longmapsto \text{Sign}(a) + j \: \text{Sign}(b)
\end{split}
\end{equation}
\end{itemize}
\end{defn}
 We equalize the analog received signal to obtain a binary output, defined using the previous mappings as follows:
\begin{equation}
\textbf{y} = \widehat{\text{ g}}(\textbf{r}) = \text{ Sign}(\operatorname{Re}(\textbf{r})) + j \text{ Sign}(\operatorname{Im}(\textbf{r}))
\end{equation}
%
%
%

\section{ML algorithms for Channel Estimation}\label{sec:ML}

The aquisition of accurate CSI is crucial for THz communication systems. Channel estimation in THz systems is a challenging task due to the tiny wavelength of THz signals, the hardware limitations eg. high speed ADCs, and difficulty of estimating large amount of parameters with small size measurements. Moreover, channel estimation needs a minimum SNR value to ensure the reliability of measurments, while the bandwidth in THz systems is  in the order of gigahertz, which presents a high thermal noise. As a results, traditional channel estimation techniques, which are based on approximative channel models may not be  sufficiant to provide an accurate channel parameters estimation. In this section we define the problem of channel estimation, and present different solutions and various ML based techniques, which learn channel parameters over any type of channel without prior assumptions.

\subsection{Problem Definition}\label{subsec:Pr-def}
 The goal of CE problem is the estimation of channel response parameters $\textbf{H}$ from the received  noisy symbols $\textbf{y}_n,  \forall n\in\mathcal{N}$, given  perfect  knowledge  of the sent pilots $\textbf{x}_n,  \forall n\in\mathcal{N}$, where $\mathcal{N}$ defines the set of indexes of the training blocks,  $N=|\mathcal{N}|$ is the training block size assumed to be equal to $M_t$. 
 We convert the complex definition of parameters into real definition, as follows:
 \begin{equation}
 y_n = [\operatorname{Re}(\textbf{y}_n), \operatorname{Im}(\textbf{y}_n)]
 \end{equation}
 
 \begin{equation}
 H = [\operatorname{Re}(\textbf{H}), \operatorname{Im}(\textbf{H})]
 \end{equation}
 
 \begin{equation}
 x_n = \begin{bmatrix}
\operatorname{Re}(\textbf{x}_n) & \operatorname{Im}(\textbf{x}_n) \\ 
-\operatorname{Im}(\textbf{x}_n) & \operatorname{Re}(\textbf{x}_n)
\end{bmatrix}
 \end{equation}
 
 \begin{equation}
 z = [\operatorname{Re}(\textbf{z}), \operatorname{Im}(\textbf{z})]
 \end{equation}
 Let $y_n^e$ be the expected equalized binary output of received signal. In order to study a variety of ML algorithms, we define more than one loss function for CE problem, considering models from the literature and proposed variations as follows:
\begin{itemize}
\item The non-linear least squares minimization:  it aims  at minimizing the error between the expected output and the calculated one.
\begin{equation}\label{eq:min1}
 \begin{split}
 [\widehat{H},\widehat{z}] = \arg{\min_{H, z}} &  \Vert y_n^e - y_n \Vert_{2},\\
 & y_n = f(Hx_n + z),  \forall n\in\mathcal{N}
 \end{split}
 \end{equation}
\begin{itemize}
\item with linear activation function \cite{chun2019deep}: In this case, we consider $f(.)$ to be the identity function, which is equivalent in neural network terminology to a perceptron with linear activation function.

\item with tanh activation function: In this case, $f(x)=tanh(x)$ is an activation function defined on $\mathbb{R}\rightarrow [-1,1]$, which gives an output close to the $Sign$ function when the input is absolutely high, while being differentiable. The differentiability of the function is important for any neural network algorithm in order to calculate the derivative.
\end{itemize}

\item The log-likelihood minimization \cite{choi2016near, myers2020low}: is designed to estimate a transformed  varaible $\textbf{X}=X^{Re} +j X^{Im} \in\mathbb{C}^{M_t \times M_r} $ corresponding to the matrix $\textbf{Hx}$. The noise component is removed from the optimization because it is considered to be constant. 
\begin{equation}\label{eq:min2}
 \begin{split}
 [\widehat{X^{Re}}] = \arg{\min_{X^{Re}}} &  \sum_{i=1}^{M_t}\sum_{j=1}^{M_r}\mathbbm{1}_{[y_{i,j}^{Re}=1]}log(\phi(X_{i,j}^{Re}/\sigma))\\& + \mathbbm{1}_{[y_{i,j}^{Re}=-1]}log(1-\phi(X_{i,j}^{Re}/\sigma))\\
 \end{split}
 \end{equation}
 where $\phi(.)$ is the cumulative distribution function of the standard normal random variable, and $\sigma$ represents its standard deviation.
\end{itemize}  
  
\subsection{Learning based Approach}
Before introducing the proposed benchmark solutions, it is essential to reformulate the minimization problem in eq. (\ref{eq:min1}) and eq. (\ref{eq:min2}) as a binary classification problem. First, we create a training dataset of $N$ data points \{$(x_n, y_n),  \forall n\in\mathcal{N}  $\}, where each data point represents one pilot. The minimization problem in eq. (\ref{eq:min1}) can be seen as a minimization of the $L2$-norm loss function. It is well known that this loss function can provide stable solutions unlike $L1$-norm, but it suffers from overfitting problem. To prevent this the coefficients from fitting perfectly to the pilot training data, we introduce a regularization term to the loss function called $L2$-regularization.\\
\begin{equation}\label{eq:min3}
 \begin{split}
 [\widehat{H},\widehat{z}] = \arg{\min_{H, z}} &  [ \mathcal{J}(H,z) ],\\
\mathcal{J}(H,z) =  & \frac{1}{N}\sum_{n=1}^{N}\Vert y_{n}^{e}- \text{tanh}(Hx_{n}+ z)\Vert _2\\ & + \lambda \Vert H \Vert_{2}
 \end{split}
 \end{equation}
 where $\lambda$ denotes a regularization coefficient. 

\subsubsection{Processing Units} 
\begin{figure}
 \centering
   \includegraphics[scale=0.3]{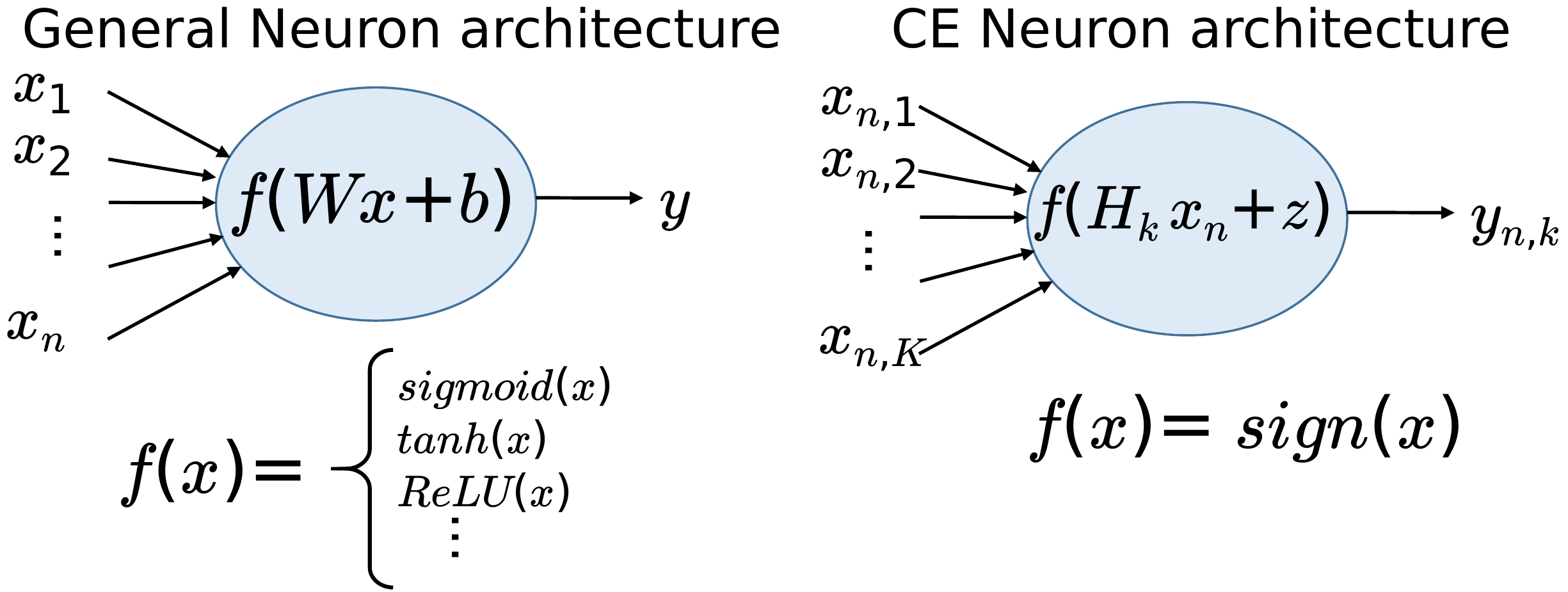}
 \caption{Neuron architecture.}
\label{fig:neuron}
\end{figure}

The basic processing unit is called artificial neuron and it computes additive and multiplicative operations over input variables, before applying an activation function. Fig. \ref{fig:neuron} illustrates both the architecture of a general artificial neuron and the instantiated version  for CE. \\ Let $W=(w_{1}, w_{2},..., w_{n})$ be the weight vector of the artificial neuron, $x=(x_{1}, x_{2}, ..., x_{n})$ is an input variables, and a bias value $b$. The process of a neuron can be represented as follows:
\begin{equation}
y = f(\sum_{i=1}^{n} w_{i} 	x_{i} + b)
\end{equation}

where y denotes the output, and $f(.):\mathbb{R}\longrightarrow \mathbb{R}$ represents an activation function.

Commonly, $f(.)$ is a non-linear function, such as  sigmoid, hyperbolic, and rectified linear function. In CE problem, the activation function of a neuron corresponds to the $sign(.)$ function applied by ADC device, which fulfills the characteristics of an activation function. 

\subsubsection{Network Components}
\begin{figure}
 \centering
   \includegraphics[scale=0.4]{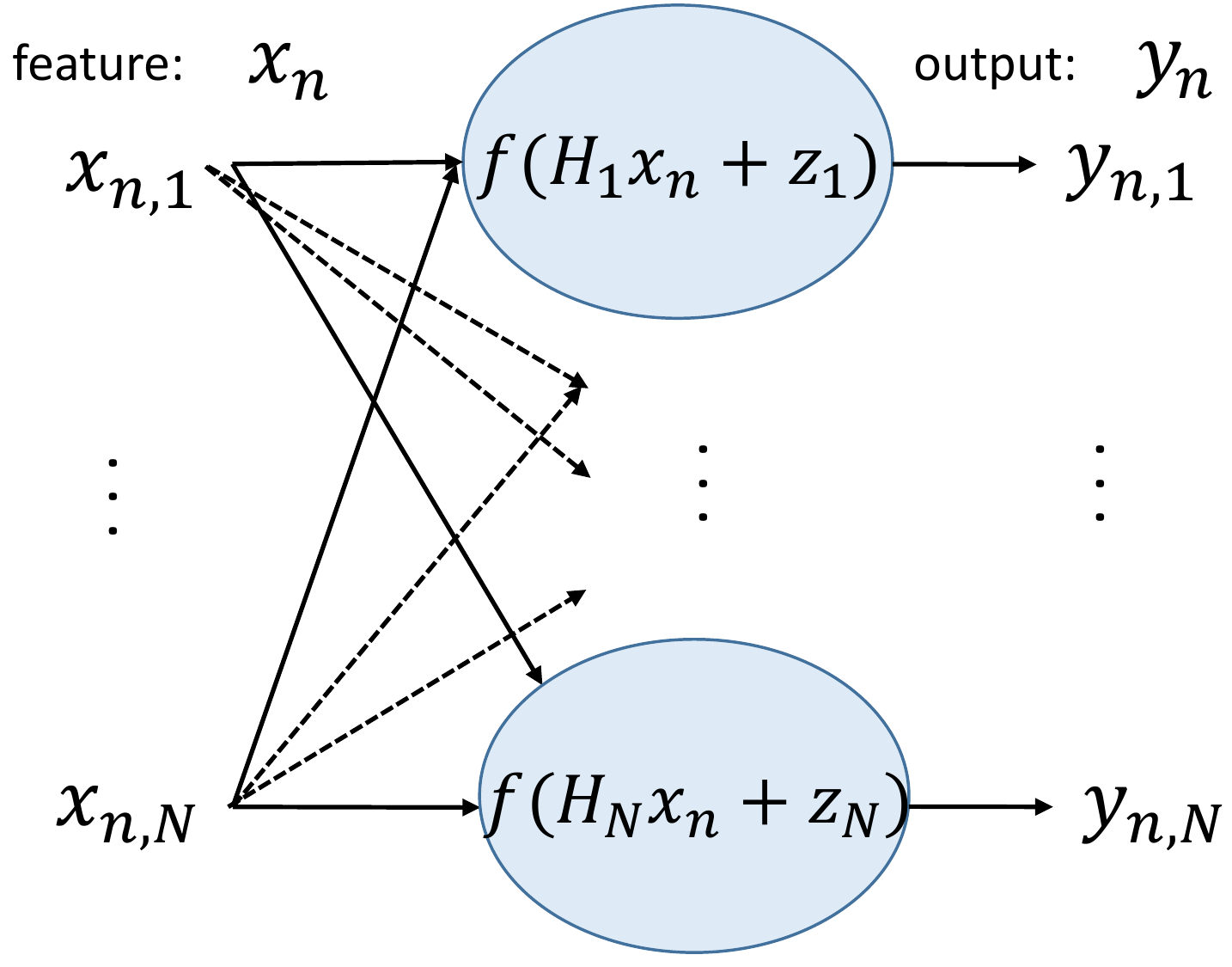}
 \caption{NN-CE architecture.}
\label{fig:NN-CE}
\vspace{-0.2 cm}
\end{figure}

Generally, neurons are grouped into layers to form a multilayered network architecture, and based on the stacking configuration of processing units, we obtain different types of neural networks, such as fully-connected, convolutional, and recurrent networks. In our case, the CE problem is seen as a basic neural network architecture, which has one hidden layer with $N$ neurons. Fig \ref{fig:NN-CE} shows the neural network based channel estimation NN-CE architecture, where the vectors $H_n=[h_{n,1}, ..., h_{nN}], \forall n \in \mathcal{N}$ represent  the channel response vector for the receive antenna $n\in \mathcal{N}$. The vector  $z=[z_{1},...,z_{N}]$ denotes the  bias parameters of NN and it can be explained as the channel noise associated to each receive antenna $n\in \mathcal{N}$.

\subsubsection{Training}
After  modeling  the NN architecture, we need to define a loss function, in  order evaluate and improve its results. The main goal of the training  is to minimize the error loss function, while adjusting the weights and bias parameters, which is eventually the same goal of CE problem. The loss function was discribed in \ref{subsec:Pr-def}.

Some optimization algorithm like Stochastic Gradient Descent (SGD) can be used to train the NN-CE, while minimizing the loss function. The optimization process updates gradually the weights representing the channel response parameters, and the bias representing the noise, in order to find the optimal solution:
\begin{equation}\label{eq:training}
 \begin{split}
 h_{i,j} = & h_{i,j} - \alpha \frac{\partial \mathcal{J}(H,z)}{\partial h_{i,j}} \\
 z_{i} = & z_{i} - \alpha \frac{\partial \mathcal{J}(H,z)}{\partial z_{i}} 
 \end{split}
 \end{equation}

 where $\alpha$ is a parameter called the learning rate, which determines how much the model learns in each step. The  backpropagation  algorithm  is  used to calculate both the weights and bias partial derivatives of the cost function.
 
 The  training  steps of  NN-CE can be described in two main steps:  (i)  the  feed-forward  pass, where the information cross all the network layers (one layer in our case) from the input layer to the output layer which contains the classification decision, i.e. the channel response, and (ii) the backpropagation pass, where the error generated by the NN-CE is calculated and propagated through the layers in the opposite direction, from the output layer to the input layer. 
 \subsection{Benchmark Algorithms}
 Variety of ML algorithms can be tested for the CE problem, where the concept of iterative learning is preserved, while modifying some details in the way we update the CSI at each step. In the previous subsection, we described how NN is modeled and used for channel estimation. Similar to it we can use Logistic Regression (LR) while changing the loss function. In our paper, we consider two other learning algorithms: Projected Gradient Ascent (PGA) and  Franke-Wolfe techniques, developed in \cite{myers2020low} for low-rank mmWave MIMO channel, in order to check its performance on THz MIMO channel, which is extremely  sensitive to weather condition, distance, and suffers from molecular attenuation. For PGA algorithm, the learning process is based on gradient ascent, a similar learning algorithm to gradient descent with a positive addition of the gradient at each update. PGA includes a  projection step at each iteration, assuming that the matrix H has lower rank $r$ than N, $r\ll N$, where  $N=M_t=M_r$. Singular Value Decomposition (SVD) and simplex projection are used to find the closest low rank matrix to the updated estimation of H. For Franke-Wolfe algorithm, an additional step is included to the basic gradient ascent learning, where we update the channel matrix differently: instead of adding a weighted gradient matrix, we compute the top singular vector of the gradient, then we soustract it from the gradient and update the channel matrix 
 \subsection{Pilot Generation Methods}
 In order to simulate the transmission process in our THz MIMO scenario, it is required to define a pilot generation method, regarding its role in TX-RX synchronization in wireless communication systems, and its importance in learning the CSI. Similar to \cite{myers2020low} we adopt two methods from the literature: DFT and ZC based sequence training. DFT and ZC sequences can be pre-determined without resorting to the knowledge  about the channel, which makes both methods suitable for our scenario.

\section{Simulation Results} \label{sec:results}

In the simulation, we consider a THz-MIMO system with $Mt=16$, $M_r=16$, and a frequency $f=0.3$ THz. We simulate the THz channel using the proposed model in section \ref{sec:SM}. The deterministic  channel model proposed in \cite{moldovan2014and} works in the range of frequencies between 0.1 and 1 THz, which explains our choice of the frequency  $f=0.3$ THz in the simulation. We consider a dry air environment, where we adopt an updated  molecular HighResolution Transmission (HITRAN) 2012 database. We consider the following gas composition of the air including their mixing ratios: $N_2$: 0.78, $O_2$: 0.21, $CO_2$: $365 \cdot 10 ^ {-6}$, $O_3$: $10 \cdot 10^{-6}$,  $CH_4$: $1.7 \cdot 10 ^{-6}$, $H_2$: $500 \cdot 10 ^{-9}$, $N_2O$: $320 \cdot 10^{-9}$, $H_2O$: 0.0096. We set the operating temperature equal to the standard temperature 296 K, and the pressure to 1 atm. 
We set the distance between the transmitter and the receiver $d=1$ meter, where this value is set to be similar to the distance used in some experiments \cite{petrov2020measurements}. The azimuth AoA and AoD are  generated from the uniform distribution in the range of $[-\frac{\pi}{2} , \frac{\pi}{2}]$. The CSI is generated using eq. (\ref{eq:h}), and the channel response at the receiver is measured as described in section  \ref{sec:SM}.  The developed THz simulator will be documented and open source, along with the generated dataset and evaluated algorithms. 

We train the proposed algorithms: Logistic Regression, NN, PGA, and Frank Wolf,  with 10 channel matrices generated with a fixed seed. The number of epochs is set to 100, which ensures the convergences based on our simulations. The stopping criterion is set the same for all learning algorithms as $\epsilon=10^{-10}$. The learning rate is intiialized with 0.01, and to $1/N_p$ , where $N_p$ is the number of pilot transmissions  for PGA algorithm. The learning rate is decreasing by 0.7 for LR and NN, when the loss function decreases, and by 0.5  with other algorithms. The previous settings are empirically chosen. The pilots used for channel measurements are generated using  ZC-based and the DFT-based training matrices. The performance of channel state information estimation algorithms are compared using the normalized mean squared error (NMSE) over all channel realizations, which is defined as the average of the Frobenius norm of  $\Vert H-\beta  \widehat{H}\Vert ^2 / \Vert H\Vert^2$, where $\beta$ is given by $\beta = \Vert \widehat{H}^\ast H \Vert_1 / \Vert \widehat{H}^\ast  \widehat{H} \Vert_1$ \cite{myers2020low}.

\begin{figure}
 \centering
   \includegraphics[scale=0.5]{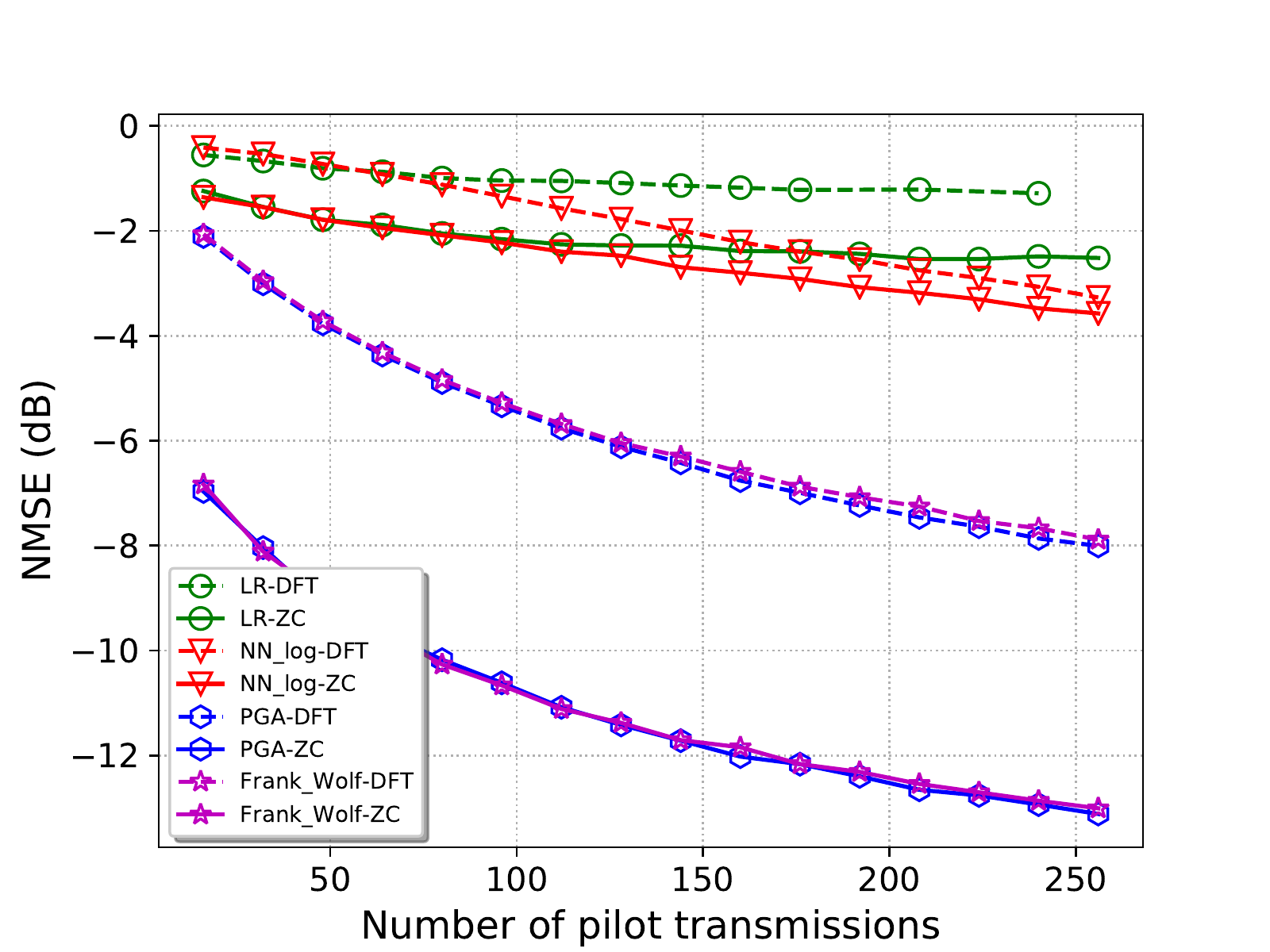}
 \caption{Performance of several learning algorithms with different pilot tranmissions number with $SNR=0$ and frequency $f=0.3 \;THz$.}
\label{fig:nmsepilots}
\end{figure}

\begin{figure}
 \centering
   \includegraphics[scale=0.5]{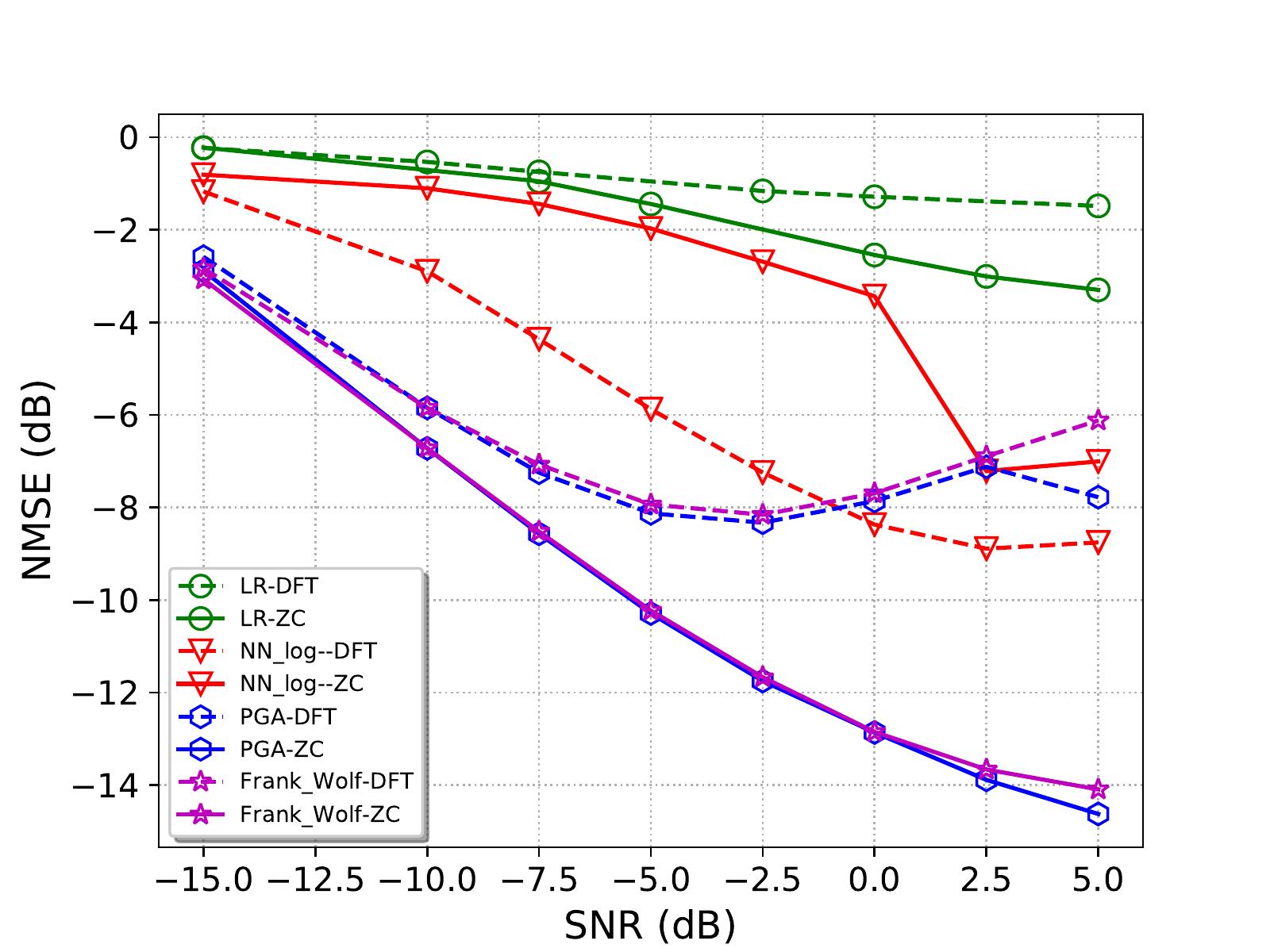}
 \caption{Performance of several learning algorithms with different $SNR$  values, with 240 pilot transmissions and frequency $f= 0.3 \;THz$.}
\label{fig:nmsesnr}
\end{figure}

Figure \ref{fig:nmsepilots} shows the NMSE of the channel estimation algorithms with different number of pilots, where the SNR is set to 0 dB, and the frequency to 0.3 THz. Each algorithm is tested using two different pilot training algorithms ZC and DFT based training. We remark that for all algorithms the error decreases when we increase the number of pilot transmissions.  The ZC-based training outperform DFT-based training for all simulations. PGA and Frank Wolf algorithms has almost the same performance with different number of pilot transmissions number, compared to LR and NN, which can be explained by the fact that those algorithms uses the knowledge about the low rank characteristics of the channel matrices, in order to reduce the dimentionality of the estimated channel information. Figure \ref{fig:nmsesnr} shows that the PGA and Frank Wolf algorithms give better results than LR and NN, for different values of SNR.  We can remark also that using DFT pilot generation is giving better results with NN than ZC method which is not the case with other algorithms. NN is using a basic network structure that can be improved in future work, regarding the potential of NN-based algorithms to learn complex structure. In THz transmission systems, it is challenging to achieve high SNR values, where $\text{SNR}=0$ is considered as a good reachable value. Both PGA  and Frank Wolf algorithms give an NMSE value around -12 dB, which is a promising result. 

\section{Conclusion} \label{sec:conclusion}
In this paper, we have modeled the channel estimation problem in THz-MIMO systems, considering a deterministic channel propagation model from the literature, and one-bit ADCs at the receiver.  We have developed several machine learning algorithms to estimate the CSI, and evaluated two different pilot training techniques: DFT and ZC based training. Our results showed a comparison between few ML algorithms that are using gradient descent as a learning optimization. PGA algorithm has given the best performance  at SNR=0 dB with NMSE of -12.8 dB. In a future work, we will compare the current algorithms to more ML and DL algorithms and other approaches from the literature, and consider studying the complexity and the hardware requirements. 
\section*{Acknowledgment}
This work was partially supported by the DFG Project Nr. JU2757/12-1, "Meteracom: Metrology for parallel THz communication channels."
\bibliographystyle{IEEEtran}
\bibliography{mybib.bib}

\end{document}